\documentclass[12pt]{article}

\usepackage{amsmath,amssymb,amsthm,epsf,epsfig}
\usepackage{amsmath,amssymb}

\begin{document}
\begin{center}
{\large {Combining The Tunneling And Anomaly Phenomena In
Deriving
the Gravitational Anomaly }}\\
\end{center}
\begin{center}
Subir Ghosh {\footnote{E-mail: sghosh@isical.ac.in}\\
Physics and Applied Mathematics Unit, Indian Statistical
Institute\\ 203 B.T.Road, Kolkata 700108, India}
\end{center}

\vskip 1.0cm {\bf{Abstract:}}\\ In the present work  we have
derived the gravitational anomaly from a fundamentally different
perspective: it emerges due to the
 tunneling of particles (in the present case fermions) across the black hole horizon. The latter effect is in
 fact the Hawking radiation. We have used the analogy of  an early
idea \cite{j1,j2} of visualizing chiral gauge anomaly as an effect
of {\it{spectral flow}} of the energy levels, from the negative
energy Dirac sea, across zero energy level in presence of gauge
interactions. This was extended to conformal anomaly in
\cite{fumita}. In the present work, we exploit the latter
formalism in black hole physics where we interpret crossing the
horizon of black hole (the zero energy level) as a spectral flow
since it is also accompanied by a change of sign in the energy of
the particle.  Furthermore, Hawking radiation induces a shrinking
of the radius of the horizon \cite{p1,p2} which reminds us of a
similar rearrangement in the Fermi level generated by the spectral
flow \cite{j1,j2,niel2}. Hence in our formulation  the negative
energy states below horizon  play a similar role as the Dirac sea.
We successfully recover the gravitational anomaly.\\

Among the various approaches to derive Hawking radiation
\cite{haw}, two have recently stood apart in popularity: (i) the
tunnelling mechanism by Parikh and Wilczek \cite{wil1} and (ii) anomaly mechanism by Robinson and Wilczek
\cite{wil2}. Furthermore, there have been numerous refinements in
the computational procedure \cite{a1}-\cite{a11}. Quite surprisingly, it
seems that these two mechanisms are disconnected both in
conceptual as well as computational sense. Indeed, it would be
very satisfying if this situation can be improved. In the present
Letter we propose a novel conjecture that the two phenomena are
complimentary to each other: the tunnelling across the
horizon (or the Hawking radiation) {\it{generates}} the gravitational anomaly. This is quite in
keeping with the spirit of \cite{p1,p2} where it is shown that
crossing of the horizon actually creates the potential barrier
through which the particle tunnels. We interpret the crossing of
the horizon by a  particle as a form of {\it{spectral flow}} across the
zero energy level that is the particle at the horizon, that
preferentially creates one type of chirality thereby generating
the anomaly. The experts in anomalies will immediately notice that
we have simply borrowed the intuitive and physical mechanism
\cite{j1,j2,b1,b2,niel2,niel3,fumita,jac} of the generation of
chiral gauge anomaly \cite{c1,c2}  where the $U(1)$ gauge interaction
induced a spectral flow across the zero energy state or the Fermi
level in Condensed Matter Physics \cite{niel2}.

One immediately notices a striking conceptual similarity between
the two phenomena: spectral flow across the Fermi level in case of
gauge anomaly \cite{j1,j2,niel2} and particle crossing the horizon
in case of black hole, especially along the lines of Parikh
\cite{p1,p2}. It has been emphasized by Nielsen and Ninomiya
\cite{j1,j2} that the Fermi level gets rearranged or shifted due
to the spectral flow so that states of one chirality are
preferentially produced. In the language of Parikh \cite{p1,p2}
also, due to Hawking radiation (or equivalently negative energy
particle falling in the black hole) the size of the  event horizon
size of the black hole gets reduced, in order to maintain energy
conservation. Furthermore, in both the instances of chiral anomaly
and gravitational (or diffeomorphism) anomaly, one gets identical
contributions coming from particle and anti-particle channels that
eventually add up to yield the final anomaly expression. The
theory becomes chiral because the ingoing modes can not influence
the physics outside the horizon and hence drops out from
contention. Indeed, this indicates that the Fermi level in
Condensed Matter Physics and event horizon in black hole physics
play analogous roles in the present instance.

Two ideas stand out in the derivation of Hawking radiation via
tunnelling \cite{wil1}: firstly the  signature of energy changes
as a particle crosses the   Schwarzschild radius or horizon (that
is induced by the metric) and secondly the Hawking radiation
survives because one partner of a particle-antiparticle pair
crosses the great divide (horizon) and  essentially becomes lost
as far as the outside world is concerned. The final Hawking
radiation gets equal contribution from both particle and
antiparticle sectors.

Let us now recall the spectral flow mechanism \cite{j1,j2,b1,b2,
niel2,niel3,fumita} (see also \cite{jac} for the simplest example)
for the chiral anomaly  \cite{c1,c2}, induced by the negative energy
Dirac sea for fermions. Even though the Dirac sea has lost some of
its ground in Quantum Field Theory, it is probably the only option
that one has in understanding the anomaly phenomenon in an
intuitive way and it should be stressed that the Dirac sea
approach \cite{j1,j2,fumita} reproduces correctly the numerical
factors in different types of  anomaly expressions as well. The
main idea here is that although classically the two chiralities of
a massless fermion (say an electron in the massless limit)
interacting with electromagnetic field are completely decoupled,
they actually become entangled through quantum effects in the
presence of Dirac sea. The gauge field induces  a (spectral) flow
of the states across the zero energy level for massless particle.
This preferentially creates one type of chirality, ({\it.{e.g.}}
right-handed antiparticles in the Dirac sea  and left-handed
particles), and  gives rise to the chiral anomaly, or
non-conservation of chiral charge, even though overall charge
conservation is maintained. Exactly similar circumstance also
prevails when one considers chiral fermions in curved background.
It was shown by Fumita \cite{fumita} that one obtains the
gravitational anomaly. In fact quantum mechanical analysis is
enough for the gauge anomaly but quantum field theory is needed
for the gravitational anomaly \cite{fumita}.

 It has been emphasized
\cite{niel3} that the whole of the infinite negative energy Dirac
sea contributes to the anomaly as that constitutes the ground
state of the system. But this is not apparent in the derivation of
the gauge anomaly via this scheme \cite{j1,j2,jac} where the
correct expression for the chiral anomaly, including numerical
factors, is reproduced by considering the contribution of the zero
(or Fermi energy) level only. This puzzle was  solved (in a
somewhat unnoticed work) by Fumita \cite{fumita} who, to our
knowledge for the first time, considered the conformal anomaly
also in the spectral flow framework. Fumita introduced a
regularization scheme - an exponential cutoff ($\Lambda $) for
large energy - to tackle the infinite negative energy sea and
obtained both the chiral and conformal anomaly by integrating over
the Dirac sea. Now it becomes clear in this unified study that
this formal procedure is really redundant for the gauge anomaly
where one is concerned with the fermion {\it{number}} current only
and the energy of the levels does not explicitly come into play.
In all the intermediate states between energies  zero and $\Lambda
$ the spectral flow just shifts the particles from one level to
another (recall the argument of putting guests in a full   hotel
having an infinite number of rooms in \cite{niel3}) and the
overall non-trivial effect can be obtained by considering the
spectral flow effect at only the zero energy level
\cite{j1,j2,jac} or at only the bottom level $\Lambda $
\cite{fumita}. On the other, when one considers the gravitational
phenomena, the {\it{energy-momentum}} current appears and to get
the ground state energy one needs to integrate the energy of an
arbitrary level over the whole (properly regularized) Dirac sea
and the whole Dirac sea explicitly contributes \cite{fumita}.

Finally we come to our conjecture that both of  the above two
phenomena, Hawking radiation in the tunnelling mechanism
\cite{wil1} and  gauge anomaly induced by the Dirac sea
\cite{jac}, are similar in nature and can be considered as
spectral flows  provided one identifies the effect of crossing
over Schwarzschild horizon in the former with crossing over the
zero energy level in the latter as both the cases involve a change
in sign of energy. Clearly the horizon (being the crossover
region) and energies close to the black hole singularity play the
zero and $\Lambda $ energy levels respectively. This is also
consistent with the accepted boundary condition of zero
energy-momentum tensor at the horizon \cite{wil1,wil2}. Clearly
this indicates that we are working in the Unruh \cite{un} or
Hartle-Hawking \cite{hh} vacuum. This is further consolidated by
the fact , as we demonstrate later, that  in our approach we
recover the covariant anomaly. The connection between covariant
boundary condition leading to covariant anomaly and the two above
mentioned vacua  are shown in \cite{kul}. Hence specification of
the vacuum is inherent in our scheme of understanding the Hawking
radiation.

In this note we will demonstrate that a proper marriage of the
above two formalisms \cite{wil1,wil2} leads to the correct form
gravitational anomaly structure. The details regarding computation
of the correct numerical factor can be obtained from the work of
Fumita \cite{fumita}. To achieve this we will apply the formalism
of deriving gravitational anomaly \cite{jac} in an external black
hole background. In comparison with the case of chiral gauge
anomaly \cite{j1,j2,jac}, we will  replace the spectral motion
across zero energy level \cite{j1,j2,jac} by the crossing of the
Schwarzschild horizon having a small width \cite{wil2}.

Before proceeding we mention a recent work \cite{kos} where it is
argued that the event horizon separates the essentially quantum
regime inside the horizon from the exterior classical region. In
\cite{bock}  also it is advocated that  the superluminal states in
the Kerr-Neumann disc can be considered as a Dirac sea of
quantized negative energy states. In fact much earlier
Chandrasekhar had speculated \cite{ch} on the significance of
negative energy states in the context of super-radiance in Dirac
waves in A Kerr-Neumann background.

We now cast the dynamics of a fermion in a curved spacetime
\cite{dar} in an analogous form of the fermion in gauge
interaction \cite{j1,j2,fumita,jac} that is suitable to study the
spectral flow. We consider the following metric,
\begin{equation}
ds^2=g_{\mu\nu}dx^\mu dx^\nu =f(r)dt^2 - f(r)^{-1}dr^2-r^2d\Omega
^2_{(d-2)}, \label{1}
\end{equation}
in $1+1$-dimensions  since near horizon the theory effectively
becomes $1+1$-dimensional \cite{f1,f2,wil2}. Next we define the
zweibein $e^a_\mu $
\begin{equation}
g_{\mu\nu }=\eta _{ab} e^a_\mu e^b_\nu ,~~ \eta _{ab} =diagonal
\{1,-1\}. \label{equation3}
\end{equation}
The flat Minkowski-space massless Dirac equation
\begin{equation}
i\gamma ^a\partial _a \psi =0,~~\{\gamma ^a,\gamma ^b\}=2\eta
^{ab} , \label{equation0}
\end{equation}
in the curved background is generalized to \cite{dar},
\begin{equation}
i\gamma ^\mu \nabla _\mu \psi =0 ,~~\{\gamma ^\mu,\gamma
^\nu\}=2g^{\mu \nu},~~\nabla _\mu =\partial _\mu
+\frac{1}{2}\sigma ^{ab}\omega _{ab;\mu}. \label{cov}
\end{equation}
In (\ref{cov}) we have defined \cite{dar} $\sigma
^{ab}=\frac{1}{4}[\gamma ^a,\gamma ^b]~, \gamma ^\mu =\gamma ^a
E^\mu _a $ with $E^\mu _a $ being the inverse of $e_\mu ^a $. For
the metric in question (\ref{1}) we have
\begin{equation}
e^a_\mu =
 \left (
\begin{array}{cc}
 \sqrt {f} & 0 \\
0 &  \frac{1}{\sqrt {f}}
\end{array}
\right ) ~~;~~E_a^\mu =
 \left (
\begin{array}{cc}
\frac{1}{\sqrt {f}}  & 0 \\
0 & \sqrt {f}.
\end{array}
\right ). \label{e}
\end{equation}
This brings the Dirac equation (\ref{cov}) to the following form,
\begin{equation}
 \left (
\begin{array}{cc}
 0 & i(\frac{\partial _t}{\sqrt {f}}+\frac{f'}{4\sqrt {f}}+\sqrt {f}\partial_r)\\
i(\frac{\partial _t}{\sqrt {f}}-\frac{f'}{4\sqrt {f}}-\sqrt
{f}\partial_r)  & 0
\end{array}
\right )\psi =0. \label{dir}
\end{equation}
To separate the chiral components  we have used the convention
\cite{jac}),
$$\gamma ^0=\sigma ^1~;~~\gamma ^1=i\sigma ^2~;~~\gamma _5=i\gamma ^0\gamma ^1=-i\sigma
^3.$$ We define the positive and negative chiral components
$\psi_{\pm}=\frac{1}{2}(1\pm i\gamma _5)\psi $
 and find
 \begin{equation}
\psi _+=\left (
\begin{array}{c}
 e^{i(-Et+pr)} \\
 0
\end{array}
\right )~~;~~\psi _-=\left (
\begin{array}{c}
 0 \\
 e^{i(-Et+pr)}
\end{array}
\right ). \label{equation1}\end{equation}
 This yields the  following set of algebraic equations,
\begin{equation}
E_-=-(fp-\frac{1}{4r_h}) - \frac{f'}{4}~~;~~E_+=(fp-\frac{1}{4r_h}) + \frac{f'}{4}, \label{chir}
\end{equation}
 for $\psi _-$ and $\psi _+$ respectively and $f'$ denotes
 $\partial_{r}f$. In the above we have taken the Schwarzschild black hole with $f=1-(r_h)/r$.
 Strictly speaking $E_{\pm}$ are the energy levels
 for constant potential but we need to generalize them for
 adiabatically varying situation. But before elaborating on this
 let us discuss in some detail the already well known derivation
 of chiral gauge anomaly in a structurally similar formalism. This
 is the central idea in our scheme.

 The flat space Dirac equation for massless
 electrodynamics is
\begin{equation}
 \gamma ^a (i\partial _a -eA_a)\psi =0,
\label{equation2}
\end{equation}
and this reduces to \cite{jac}
\begin{equation}
E_-=-p+eA_1~~;~~E_+=p-eA_1 \label{an1}
\end{equation}
for $\psi _+$ and $\psi _-$ respectively and $\dot A_1=-E$, the
electric field. (The analogous set of equations in our case are
(\ref{cov}) and (\ref{chir}) respectively. Before proceeding
further I should mention how to interpret (\ref{an1}) and
(\ref{chir}). The notation $E_{\pm}$ are not to be understood as
energy eigenvalues. In fact they would have been energy levels for
constant potentials. These relations are in a classical setting
and they simply show in a heuristic way how a change of potential
with time induces the spectral flow. A detailed derivation of the
anomaly equation exploiting classical physics is provided in
\cite{ws} where the underlying quantum nature of the field theory
manifests itself only through the negative energy Dirac sea. In
fact the true energy levels are not at all needed for the present
derivation (see for example \cite{j1,j2,niel2,jac,fumita}).

More specifically, in $A_0 =0$ gauge, the only non-zero gauge
invariant quantity is the electric field $E=\partial _t A_1$ and
this time variation is introduced in $A_1$ (for the chiral gauge
anomaly case in \cite{jac} by Jackiw)  as an {\it{adiabatic}}
change $A_1 \rightarrow A_1 +\delta A_1$.  For an adiabatic change
the structure of the energy spectrum is not changed but only the
explicit value of the energy level gets altered. Hence the
particle states are quasi-stationary in nature.

  A change in the gauge field
induces the spectral flow across the zero energy level which
signals the creation of one type of chirality. This is directly
identified as the rate of change of fermion number of that
chirality. This can be visualized from the Figures 1 and 2
\cite{jac}. In Figures 1 and 2 we plot the dispersions (\ref{an1})
for $A_1=0$ and  for a non-zero $A_1$ with $\delta A_1 >0$
respectively. The acute and obtuse angled branches refer to left
handed and right handed chirality states respectively. The filled
states are shaded circles whereas the empty states are clear
circles. The ground state for $A_1=0$ is shown in Figure 1 with
the filled up negative energy sea. In Figure 2 the movement of the
states (or the spectral flow) across the zero energy level for an
adiabatic change in $A_1$ with $\delta A_1 >0$ is shown by the
arrows and both branches contribute equally to the creation of
right handed chirality.  For a quantization length $L$ with
density of states per length $L$ being $L/2\pi$, the rate of
production of RH particle is
\begin{equation}
 \dot N_{RH}=L^{-1}(L/2\pi)\dot \omega_{FS}=(e/2\pi)E,
\label{eq0}
\end{equation}
where $\omega_{FS}$ is the energy at the Fermi surface. Using the
massless dispersion relation for energy $\omega$ and momentum $P$
and Newtonian dynamics one connects $\dot\omega =\dot P=eE$. To
this one adds the identical contribution coming from the Left
Handed sector and thus obtains the axial anomaly as $\dot
N_{RH}+\dot N_{LH} =(e/\pi)E$. This is the standard "Quantum
Mechanics" derivation \cite{j1,j2,jac} where only the effect at
the zero energy level appears.

In the field theoretic analysis of Fumita \cite{fumita} one
studies the classical conservation law
\begin{equation}
 \partial_{\mu}j^{\mu}_{5[-\epsilon(\Pi )]}=0~;~~j^{\mu}_{5}=\bar\psi\gamma^{\mu}\gamma_{5}\psi,
\label{f1}
\end{equation}
where $j^{\mu}_{5\omega(\Pi  )}$ denotes the axial current for
energy $\omega =\Pi =P-eA_1$. For the anomaly one takes into
account the whole Dirac sea by integrating the regularized
current:
\begin{equation}
 L\int \frac{d\Pi}{2\pi}j^{\mu}_{5[-\epsilon(\Pi )]}exp(-\frac{\Pi ^2}{\Lambda ^2}).
\label{f21}
\end{equation}
Using the spectral flow arguments as before for the energy level
$\Lambda $ one recovers \cite{fumita} the anomalous Ward identity
\begin{equation}
 \partial_{\mu}j^{\mu}_{5}=\frac{eE}{\pi}.
\label{eq}
\end{equation}
As we pointed out before, in actuality $j^{\mu}_{5}$ is the
fermion number current and does not contain any $\Pi $ and thus
one can get the correct answer without considering  the whole
Dirac sea or a regularized current \cite{j1,j2}.

 Now we come to
 the
 gravitational
 anomaly. In Figure 3 we have tried a similar picturization (as the gauge anomaly explained previously)
 corresponding to (\ref{chir}). The first and second diagrams in Figure 3 depict the spectra for the particle (\ref{chir}) outside and inside the horizon respectively. Upper halfs of both the diagrams are only relevant that show a preferential change in the chirality content after crossing inside the horizon.  Clearly $f'$
 in
 (\ref{chir})
 plays the role
 of the gauge
 field. If we
 apply our
 mechanism in a
 thin slab
 straddling the
 horizon
 \cite{wil2}
 the change in
 the
 gravitational
 field
 witnessed by a
 particle as it
 crosses the
 horizon will
 be $\epsilon
 f''$ with
 $\epsilon$
 being the
 thickness of
 the slab.
 Taking a time
 derivative
 yields
 \begin{equation}
 \epsilon \dot
 f''=\epsilon
 f'''\dot
 r=\epsilon
 f'''f.
 \label{an}
 \end{equation}
 In the last
 step we have
 considered the
 radial null
 geodesics and  have used
 $\dot r=f$.
 Considering
 the above
 equation
 (\ref{an}) per
 unit length,
 one gets the
 correct
 structure of
 the covariant
 form of
 gravitational
 anomaly
 \cite{d1}-\cite{d5}, \cite{wil2}
 as $f'''f$.
 Conceptually,
 the most
 direct usage
 of black hole
 physics in the
 present paper
 lies in the
 part of
 exploiting the
 radial null
 geodesic
 condition in
 replacing
 $\dot r$ by
 $f$. We are
 still not
 finished since
 the correct
 numerical
 factor in the
 anomaly has to
 be
 ascertained.
 In fact, in
 analogy with
 the fermion
 case, what we
 have obtained
 is the anomaly
 in the fermion
 number and not
 the
 gravitational
 anomaly that
 deals with the
 energy-momentum
 current
 (schematically
 $\sim
 \bar\psi\gamma^{\mu}\partial^{\nu}\gamma_{5}\psi
 \sim
 k^{\nu}\bar\psi\gamma^{\mu}\gamma_{5}\psi$).
 This requires
 a careful
 analysis since
 one has to
 consider the
 energy
 contribution
 due to the
 above for each
 negative
 energy level
 and then
 integrate over
 the whole
 Dirac sea with
 proper
 regularization.
 However, it is
 important to
 note that the
 above analytic
 expression
 $f'''f$ for
 each level is
 independent of
 the level
 momentum (or
 energy) and
 will come out
 of the energy
 integral of
 the Dirac sea
 which will
 finally
 reproduce only
 the numerical
 factor.  But
 precisely this
 effect has
 already been
 computed by
 Fumita
 \cite{fumita}
 and we simply
 borrow his
 result. Fumita
 in
 \cite{fumita}
 discusses both
 the boson and
 fermion
 sectors in
 conformal
 gauge from the
 spectral flow
 point of view.
 For the
 fermions, the
 regularized
 vacuum energy
 functional
 \cite{fumita}
 \begin{equation}
 E[\phi
 (x)]=-L\int^{\Lambda}_{0}\frac{dk}{2\pi}\frac{\mid
 k
 \mid}{2}<\psi_{k},exp(-\frac{\Delta
 ^{(1/2)}}{\Lambda
 ^2}\psi_{k}>_{spinor},
 \label{f11}
 \end{equation}
has to be calculated where the positive energy wavefunction of
momentum $k$ is $\psi_{k}=exp(ikx-i\mid k\mid t)/\sqrt{L}$ which
is right (left) handed for $k>0~(k<0)$. For Majorana fermions with
conformal weight $1/2$ the general expression for the one
dimensional Laplacian $\Delta =-\mid g^{11}(x)\mid \nabla_1
\nabla_1$ where $$\Delta
^{(j)}=-e^{-\phi}(\partial_{1}-\frac{j+1}{2}\partial_{1}\phi
)(\partial_{1}-\frac{j}{2}\partial_{1}\phi )$$ reduces to
$\Delta^{(1/2)}$. Also the limits of integration relates to the
horizon and the bottom of the Dirac sea, as explained earlier. The
Liouville action is correctly reproduced (for details see
\cite{fumita}),
\begin{equation}
 E[\phi (x)]=\int dx [-\frac{\Lambda ^2}{4\pi}e^{\phi}-\frac{1/2}{96\pi}(\partial_{1}\phi )^2+\frac{1}{48\pi}\partial^{2}_{1}\phi +O(\frac{1}{\Lambda ^2})].
\label{f2}
\end{equation}
From this using the definition
$T_{\mu\nu}=\frac{2}{\sqrt{-g}}\frac{\delta E[\phi ]}{\delta
g^{\mu\nu}}$ for the energy momentum tensor one recovers the
correct form of conformal (or Weyl) anomaly,
\begin{equation}
 T^{\mu}_{\mu}=-\frac{f''}{48\pi}.
\label{f3}
\end{equation}
To bring the anomaly in this form, we had to use the
parameterization $exp(2\phi )=f$ and subsequently use
$(\partial/\partial_{x}=f(\partial/\partial_{r})$ to go to the
Schwarzschild gauge (from the conformal gauge) \cite{e1,e2,e3}.
From the trace anomaly itself Christensen and Fulling \cite{f1,f2}
have shown how to derive the Hawking radiation. Hence, comparing
with our previously obtained anomaly expression (\ref{an}), we
find that the correct numerical factor is $1/(96\pi )$
\cite{d1}-\cite{d5}.

Finally, Hawking radiation can consist of both fermions as well as
bosons. For the latter there is a consistent formulation
\cite{g1,g2} with a Dirac like negative energy sea for the vacuum.
Also in the paper by Fumita \cite{fumita} (whose formalism we have
adopted) both fermions and bosons are considered. Hence it is
expected the same excercize can be repeated for the bosons as
well.

To conclude, we have achieved what we set out to prove: the
understanding of Hawking radiation as a tunnelling phenomenon
\cite{wil1} and from gravitational anomaly framework \cite{wil2}
are not to be considered as isolated and distinct formalisms. On
the other hand, as we show here, crossing the horizon (or the
tunnelling picture) generates the anomaly when one views it as a
spectral flow of energy levels from one signature to another as it
crosses the zero level or horizon.
 We stress that in our derivation we have used  boundary conditions that are consistent with the
 accepted one that the energy momentum flux on the horizon is zero (the Unruh or Hartle-Hawking vacua). The fact
 that we are considering
 black hole physics (instead of the conventional anomaly as in \cite{fumita}) comes out strongly in our
 use of the null radial null geodesics condition on the horizon (in the derivation of (\ref{an})). Lastly we
 point out that the  numerical
 factor for the anomaly also comes correctly (using the analysis of \cite{fumita}) thereby demonstrating
 consistency and completeness of the
 spectral flow framework.\\
{\it{Acknowledgements}}: Discussions with Rabin Banerjee, Bibhas
Ranjan Majhi and Sujoy Modak are gratefully acknowledged. Also I thank Satya Prakash Ojha for his help in drawing the
figures.

\newpage
\begin{figure}
\includegraphics[height=7in]{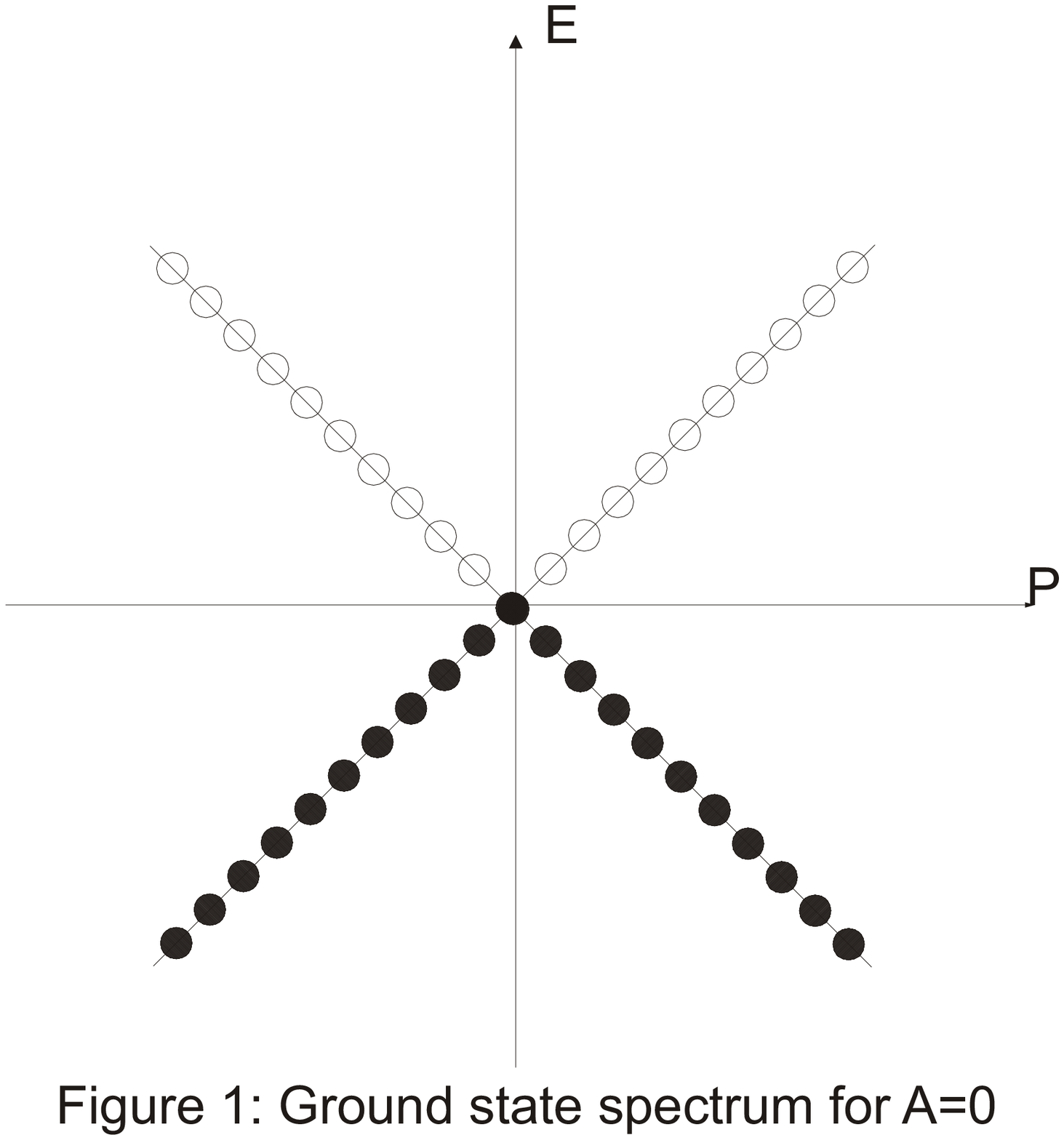}\\
\end{figure}

\begin{figure}
\includegraphics[height=7in]{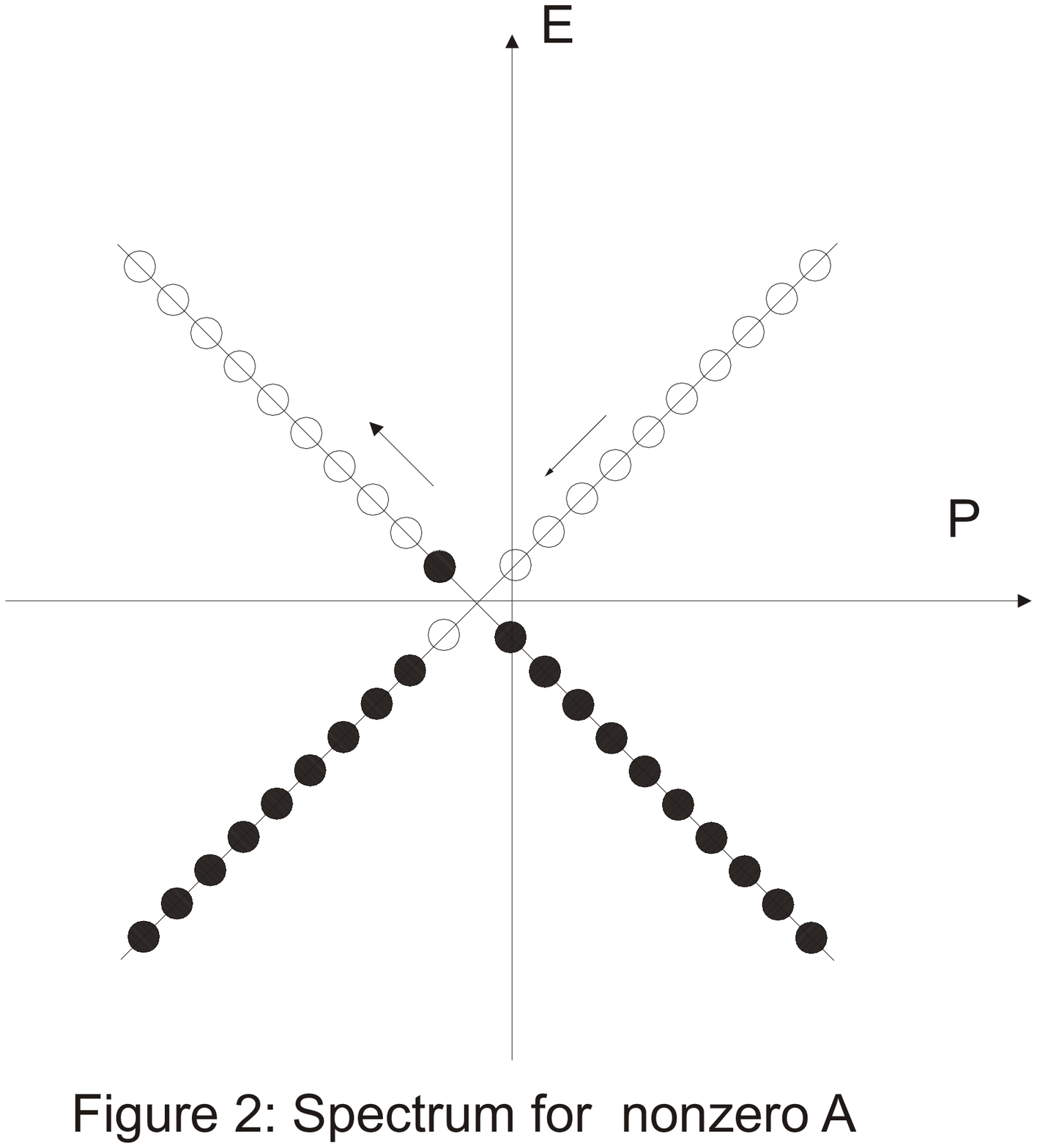}\\
\end{figure}

\begin{figure}
\includegraphics[height=9in]{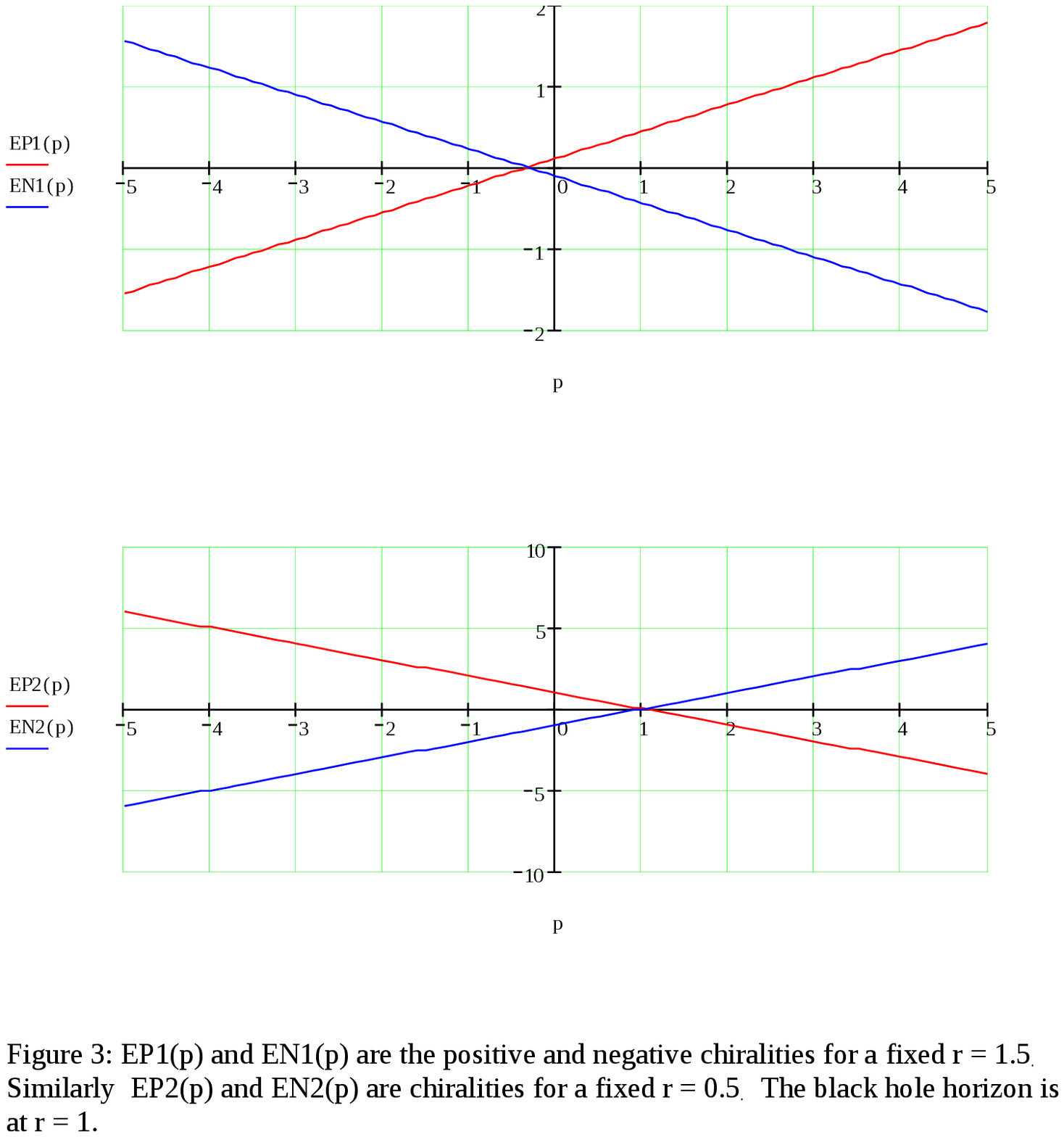}\\
\end{figure}


\vskip
.5cm

\begin{thebibliography}{99}
\bibitem{haw}S.Hawking, Commun.Math.Phys. 43 (1975) 199.
\bibitem{wil1}M.K.Parikh and F.Wilczek, Phys.Rev.Lett. 85 (2000) 5042 [arxiv:hep-th/9907001].
\bibitem{wil2}S.P.Robinson and F.Wilczek, Phys.Rev.Lett. 95 (2005) 011303 [arxiv:gr-qc/0502074].
\bibitem{a1}S.Iso, H.Umetsu and F.Wilczek, Phys.Rev.Lett. 96 (2006) 151302 [arxiv:hep-th/0602146].
 \bibitem{a2}K.Murata and J.Soda, Phys.Rev. D74 (2006)044018 [arxiv:hep-th/0606069].
\bibitem{a3} E.Vagenas and S.Das, JHEP 0704 (2007) 068 [arxiv:hep-th/0606077].
\bibitem{a4} M.R.Setare, Eur.Phys.J. C49 (2007) 865 [arxiv:hep-th/0608080].
\bibitem{a5} Q.Q.Jiang and S.Q.Wu, Phys.Lett. B647 (2007)200 [arxiv:hep-th/0701002].
\bibitem{a6} H.Shin and W.Kim, arXiv:0705.0265.
\bibitem{a7} S.Das, S.P.Robinson and E.C.Vagenas, arXiv: 0705.2233.
\bibitem{a8} R.Banerjee and S.Kulkarni, Phys. Rev. D 77, 024018 (2008)
[arxiv:0707.2449].
\bibitem{a9} R.Banerjee and B.R.Majhi, Phys.Lett.B662 62 (2008) [arxiv:0801.0200].
\bibitem{a10} L. Bonora, M. Cvitan, JHEP05(2008)071
(arXiv:0804.0198).
\bibitem{a11} L.Bonora, M.Cvitan, S.Pallua and I.Smolic,
arXiv:0808.2360.
\bibitem{p1}M.K.Parikh, Int.J.Mod.Phys. D13 (2004) 2351 [hep-th/0405160].
\bibitem{p2}M.K.Parikh, hep-th/0402166.
\bibitem{j1}H.B.Nielsen and M.Ninomiya, Nucl.Phys. B185 (1981)20.
\bibitem{j2}H.B.Nielsen and M.Ninomiya, Phys.Lett. 105B (1981) 219.
\bibitem{b1}G.'Hooft, Phys.Rev.Lett. 37 (1976) 8.
\bibitem{b2} A.S.Blaer, N.H.Christ and J.-F. Tang, Phys.Rev.Lett. 47 (1981) 1364.
\bibitem{niel2}H.B.Nielsen and M.Ninomiya, Phys.Lett. 130B (1983)389.
\bibitem{niel3}H.B.Nielsen, Talk at Trieste Conference on "Topological Methods in Quantum Field Theory", Trieste, 1990.
\bibitem{fumita}N.Fumita, Int.J.Mod.Phys. A10 (1995) 2579 [arxiv:hep-th/9406037].
\bibitem{jac}R.Jackiw, {\it{Dirac Prize Lecture, Trieste 1999}} [arxiv:hep-th/9903255].
\bibitem{c1} S.Adler, Phys.Rev. 177 (1969) 2426.
\bibitem{c2} J.S.Bell and R.Jackiw, Nuovo Cimento 60A (1969) 4.
\bibitem{un}W.G.Unruh, Phys.Rev.D14 (1976) 870.
\bibitem{hh} J.B.Hartle and S.W.Hawking, Phys.Rev. D13 (10976) 2188.
\bibitem{kul} R.Banerjee and S.Kulkarni,  arXiv:0810.5683.
\bibitem{kos}B.P.Kosyakov, Found.Phys. 38 (2008)678.
\bibitem{bock}R.Bock, arXiv: 0806.1674.
\bibitem{ch}S.Chandrasekhar, {\it{The Mathematical Theory of Black Holes}}, Johm Wiley: New York (1984).
\bibitem{dar} S.K.Moayedi and F.Darabi, Phys.Lett. A322 (2004) 173 [arXiv:gr-qc/0208053].
\bibitem{ws}A Widom and Y.Srivastava, Am.J.Phys. 56 (9) (1988)824.
\bibitem{d1} L.Alvarez-Gaume and E.Witten, Nucl.Phys. B234 (1983) 269.
\bibitem{d2} W.A.Bardeen and B.Zumino, Nucl.Phys. B244 (1984) 421.
\bibitem{d3} K.Fujikawa, Z.Phys. C28 (1985) 289.
\bibitem{d4}R.A.Bertlmann and E.Kohlprath, Ann.Phys. 288 (2001) 137
[arxiv:hep-th/0011067].
\bibitem{d5} See also R.A.Bertlmann, {\it{Anomalies in
Quantum Field Theory}}, Oxford Sciences, Oxford, 2004.
\bibitem{e1}  D.Grumiller, W.Kummer and  D.V.Vassilevich, Phys.Rept. 369 (2002) 327 [hep-th/0204253v9].
\bibitem{e2} D.R.Karakhanyan, R.P.Manvelyan and R.L.Mkrtchyan, Phys.Lett.B 329 (1994) 185.
\bibitem{e3} R.Jackiw, arXiv:hep-th/9501016.
\bibitem{f1} S.Christensen and S.Fulling, Phys.Rev. D15 (1977) 2088.
\bibitem{f2}  R.Balbinot and A.Fabbri, Phys.Rev. D59 (1999) 044031 [hep-th/9807123v1].
\bibitem{g1} Y.Habara, Y.Nagatani, H.B.Nielsen and M.Ninomiya, Int.J.Mod.Phys.A23 (2008) 2771 (arXiv:hep-th/0607182).
\bibitem{g2} Y.Habara, Y.Nagatani, H.B.Nielsen and M.Ninomiya, Int.J.Mod.Phys.A23 (2008) 2733 (arXiv:hep-th/0603242).


\end{thebibliography}
\end{document}